\documentclass[conference]{IEEEtran}
\IEEEoverridecommandlockouts
\usepackage{cite}
\usepackage{amsmath,amssymb,amsfonts}
\usepackage{algorithmic}
\usepackage{graphicx}
\usepackage{textcomp}
\usepackage{xcolor}
\usepackage{cite}
\usepackage{CJK}
\usepackage{svg}
\usepackage[caption=false,font=normalsize,labelfont=sf,textfont=sf]{subfig}
\usepackage[colorlinks=true, linkcolor=blue, citecolor=blue,urlcolor=black]{hyperref}
\usepackage{cleveref}
\usepackage{booktabs}
\usepackage{steinmetz}
\usepackage{url}
\usepackage{makecell}

\def\BibTeX{{\rm B\kern-.05em{\sc i\kern-.025em b}\kern-.08em
    T\kern-.1667em\lower.7ex\hbox{E}\kern-.125emX}}
    
\begin{document}

\title{Dual-Hop Joint Visible Light and Backscatter Communication Relaying under Finite Blocklength}

\author{
\IEEEauthorblockN{
Boxuan Xie\IEEEauthorrefmark{1},
Lauri Mela\IEEEauthorrefmark{1},
Alexis A. Dowhuszko\IEEEauthorrefmark{1},
Jiacheng Wang\IEEEauthorrefmark{2},
Kalle Ruttik\IEEEauthorrefmark{1},
Riku Jäntti\IEEEauthorrefmark{1}
}
\IEEEauthorblockA{\IEEEauthorrefmark{1}Department of Information and Communications Engineering, Aalto University, 02150 Espoo, Finland}
\IEEEauthorblockA{\IEEEauthorrefmark{2}College of Computing and Data Science, Nanyang Technological University, Singapore 639798}
\thanks{This work is partially supported by the funding through AMBIENT-6G project under the European Union’s Horizon Europe research and innovation programme
(Grant Agreement No 101192113), and the WALLPAPER project from Research Council of Finland (Grant Agreement No 352912). }
\vspace{-20pt}
}

\maketitle

\begin{abstract}
This paper investigates a dual-hop joint visible light communication (VLC) and backscatter communication (BC) relaying framework under the finite blocklength (FBL) constraint, aiming at energy-neutral Ambient Internet of Things (A-IoT) deployments. 
In the proposed system, indoor LED access points are used to simultaneously provide illumination and transmit information over light to a backscatter device (BD), which harvests optical energy and backscatters the received messages to user equipments (UEs) equipped with radio frequency (RF) front ends. 
This forwarding of the information from VLC to RF channels is implemented without the need for carrier synthesizers and power amplifiers at the IoT node. 
By modeling the end-to-end communication link with short-packet IoT traffic and realistic levels of interference between adjacent VLC coverage areas, we analyze the outage performance and achievable data rate of the proposed system. 
Simulation results demonstrate that key factors, such as placement and orientation of the BD, as well as the selected code rate of the system affect reliability and data rate that can be achieved for communication purposes. The insights gained from this study pave the way for ambient power-enabled IoT solutions and future hybrid VLC/RF network designs.
\end{abstract}


\section{Introduction}
With the advent of the sixth generation (6G) era, visible light communication~(VLC) has emerged as a key enabling technology for the energy-efficient connectivity of Internet of Things (IoT).
VLC leverages the ubiquitous light-emitting diode~(LED) infrastructure for both indoor illumination and wireless data transmission, thereby addressing the dual needs of lighting and connectivity. 
VLC-enabled IoT solutions are attractive because of their advantages in energy consumption and spectrum utilization~\cite{haas2019survey}. 
Recent advances propose hybrid indoor wireless networks that combine VLC with radio frequency (RF) communications, enhanced by simultaneous lightwave information and power transfer~(SLIPT)~\cite{ding2018slipt,xiong2024hybrid2}.
This method harnesses the capability of VLC to convey both information and power to IoT devices, supporting battery-free operations ranging from smart homes to industrial automation.

Despite its promising prospects, the deployment of VLC-enabled IoT systems faces several challenges. Existing solutions in this regard typically require complex reception and relaying mechanisms, often relying on sensitive photodetectors and retroreflectors to forward VLC signals from LED access points~(APs) to the target receiver. 
An alternative approach involves converting the incoming VLC signals into outgoing RF signals, which are forwarded using an intermediate IoT node. 
Several studies have proposed VLC-RF relaying architectures where the VLC signal is received, decoded, and then encoded again for its transmission in the RF domain such that conventional RF receivers can process the data~\cite{letaief2021hybrid, raouf2024outage}. 
However, these methods assume the presence of power amplifiers and active RF transmitters in the relay nodes, which increases hardware complexity and implementation costs. 
Furthermore, existing works often overlook the impact of the finite blocklength~(FBL) traffic, a critical factor in short-packet communications that is commonly considered in practical IoT networks~\cite{verdu2010fbl, shi2025short}. 
This oversight underscores the need for system models that fully account for the non-asymptotic performance of IoT applications.

Recent advances in Ambient IoT within the 3rd Generation Partnership Project (3GPP) have underscored the potential of ultra-low-power backscatter communication (BC) as a foundational technology for 6G cellular IoT networks~\cite{3gpp38848, sheikh2021cellular}.
In BC scenarios, backscatter devices (BDs) rely on existing RF signals for carrier modulation, eliminating the need for energy-intensive RF components, such as local oscillators and power amplifiers~\cite{duan2020ambient}. 
In this context, integrating VLC with BC offers a promising alternative to hybrid VLC/RF systems for VLC-RF relaying. Early studies have shown that low-complexity BDs can relay information from LED APs to RF receivers without the strict hardware requirements of existing active relays~\cite{xie2024vlc,xie2024light, varshney2024lifibc}. However, comprehensive system modeling and rigorous performance analysis of such systems remain unexplored.
This paper addresses these gaps by investigating a dual-hop VLC-BC relaying scheme under FBL constraints in an indoor environment with multiple LED APs. 
Key contributions of this paper are outlined as follows:
\begin{itemize}
    \item We investigate a dual-hop VLC-BC relaying scheme for indoor hybrid VLC/RF systems, where a BD receives VLC messages and relays them to user equipments (UEs) equipped with RF receivers. This is achieved by modulating the photovoltaic-converted message with ambient RF carrier waves emitted by an RF source (RFS).
    \item We develop a comprehensive system model incorporating the effect of FBL traffic, which represents a critical factor for short-packet IoT transmissions. 
    We also derive expressions for the outage probability and achievable data rate of the system, quantifying the system performance under the practical FBL constraint.
    \item We analyze the effect of placement and orientation of the BD on overall link quality under illumination from multiple LED APs. We adopt the 3GPP Indoor Hotspot channel model to assess the communication performance under different indoor radio propagation conditions.
\end{itemize}
The rest of the paper is organized as follows. Section II proposes the system model. 
Section III analyzes communication outage performance of the system.
Section IV presents and evaluates the simulation results.
Finally, Section V draws conclusions and discusses future work.
\section{System Model}
\begin{figure}
\centering
\includegraphics[width=0.98\columnwidth]{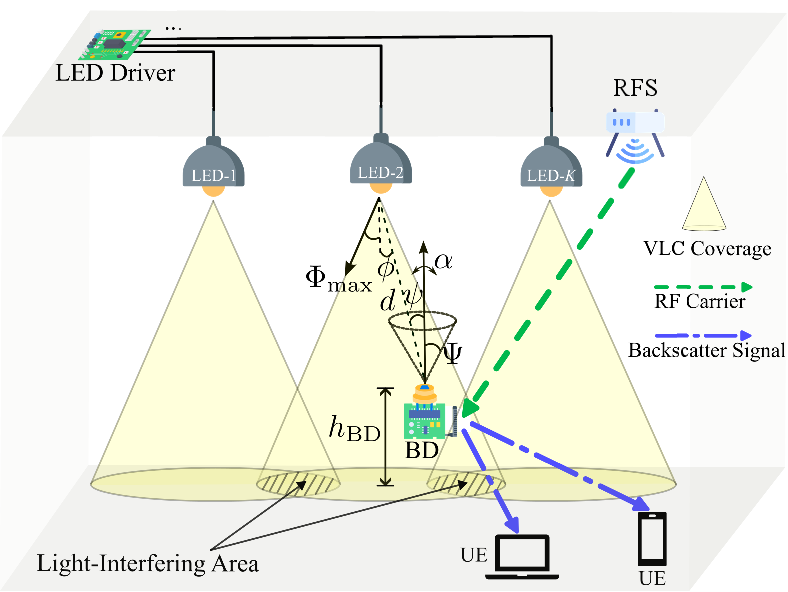}
\vspace{-5pt}
\caption{System model.}
\label{fig:system_model_general}
\vspace{-10pt}
\end{figure}

\begin{figure}
\centering
\includegraphics[width=0.98\columnwidth]{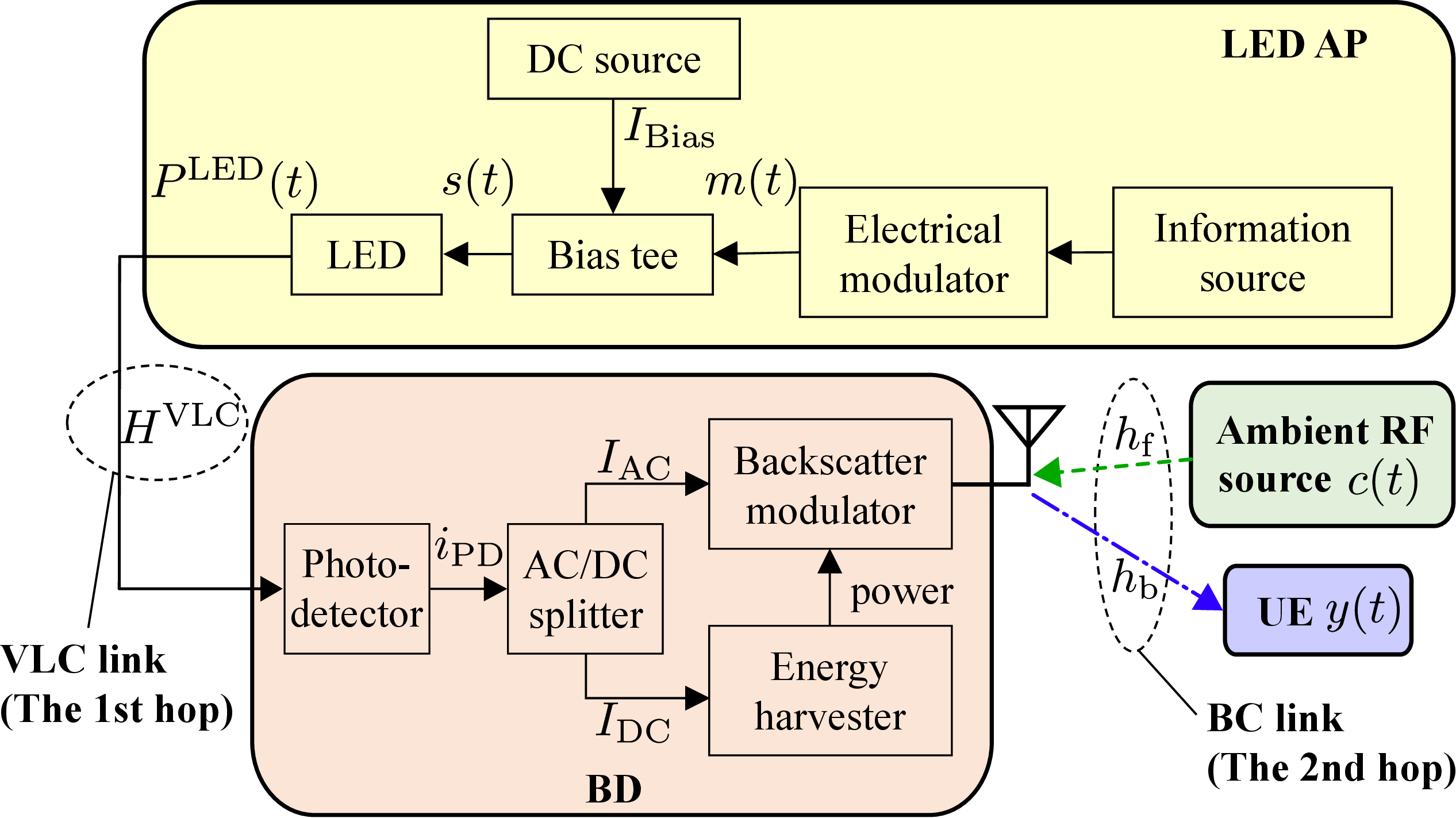}
\caption{Schematic of joint VLC-BC relaying framework.}
\label{fig:bd_scheme}
\vspace{-15pt}
\end{figure}

The proposed dual-hop VLC-BC relaying system is illustrated in Fig.~\ref{fig:system_model_general} and the schematic diagram is presented in Fig.~\ref{fig:bd_scheme}.
In the first hop (VLC link), 
LED APs transmit VLC signals using an intensity modulation/direct detection (IM/DD) scheme to carry modulated messages, which are received by a BD within their coverage areas.
The BD generally consists of a photodetector, an energy harvester, a backscatter modulator, and an RF antenna~\cite{xie2024vlc,xie2024light}.
For the proposed system targeting low-data-rate communications, this paper considers the use of photovoltaic cells as the photodetector.
The alternating current/direct current (AC/DC) splitter separates the photovoltaic-converted signals into electrical AC and DC components. The AC component, which carries the VLC messages, is used to control the backscatter modulation, whereas the DC component supplies energy to the BD.
In the second hop (BC link), the BD modulates ambient RF carrier waves, such as those emitted by an indoor RFS which could typically be, e.g., a WiFi AP. 
The modulation is controlled by the AC component which alternates the reflection coefficients of the BD.
The modulated signal is then backscattered and captured by UEs equipped with RF front ends.
Subsequently, the VLC and BC links are modeled, respectively.
\subsection{VLC Channel Model (the first hop)}
The downlink VLC transmission model from LED APs to the BD is considered for SLIPT, serving both communication and energy harvesting purposes.
\subsubsection{Waveform Design and Communication}
In the downlink direction of the VLC transmission, each LED AP is assigned a unique identifier and transmits its individual VLC signal using a time-division multiplexing (TDM) method.
On-off keying (OOK) baseband modulation is used to generate the AC current that modulates the intensity of the light emitted by the LED, while a bias tee is employed to add a DC bias to the modulating signal.
Let $m(t)$ denote the modulating signal. Then, the electrical signal that drives the $k$-th LED AP can be written as $s_{k}(t) = m_{k}(t) + I_{\textrm{Bias}}$. Thus, the transmitted optical signal can be expressed by~\cite{ding2018slipt}
\begin{equation}
    P_{k}^{\textrm{LED}}(t) =  \eta_{\textrm{E-O}} \left( m_{k}(t) + I_{\textrm{Bias}} \right),
\end{equation} 
where $\eta_{\textrm{E-O}}$ is the LED power electric-optical conversion factor, and $I_{\textrm{Bias}}$ represents the DC current driving the LED.
Since a practical LED transmitter operates within a limited linear region, the peak amplitude $s_{\textrm{max}}$ of $s_{k}(t)$ should be bounded to avoid clipping. Specifically, 
$s_{\textrm{max}} \leq \textrm{min} \left( I_{\textrm{Bias}} - I_{\textrm{min}}, I_{\textrm{max}} - I_{\textrm{Bias}} \right),$
where $I_{\textrm{max}}$ and $I_{\textrm{min}}$ denote the maximum and minimum driving current, respectively. 

To characterize the VLC channel, a direct illumination model is adopted, based on the assumption that the direct optical signal is significantly stronger than any reflected signals~\cite{alexis2024planning}. 
Phosphor-coated LEDs are considered in this paper, which are assumed to emit light following Lambertian radiation patterns, as illustrated in Fig.~\ref{fig:system_model_general}. Let $K \geq 1$ represent the total number of LED APs, including $\textrm{LED-1}$, $\textrm{LED-2}, \ldots, \textrm{LED-}K$, each of them maintaining a line-of-sight (LoS) link to the BD within its coverage area.
For simplicity, this work assumes that the BD and its photodetector share the same position.
The DC gain of the optical channel between $\textrm{LED-}k$ and BD-equipped photodetector can be expressed as~\cite{alexis2024planning}
\begin{equation}
H_{k}^{\textrm{VLC}} =
\begin{cases}
\frac{(\nu+1)A_{\textrm{PD}}}{2\pi d_k^2} \cos^\nu(\phi_k) \cos(\psi_k), & |\psi_k| \leq \Psi, \\
0, & |\psi_k| > \Psi,
\end{cases}
\end{equation}
where $d_k$ denotes the Euclidean distance between $\textrm{LED-}k$ and the BD, while $\nu = -1 / \log_2 [\cos(\Phi_{\textrm{max}})]$ refers to the Lambertian index of the LEDs. 
The parameter $\Phi_{\textrm{max}}$ indicates the semi-angle at half-power of the LED luminaire, where $\Psi$ defines the field-of-view (FoV) semi-angle of the BD-equipped photodetector with an active area of $A_{\textrm{PD}}$. Furthermore, $\phi_k \geq 0$ and $\psi_k$ are the irradiance and incidence angles of the LoS link between $\textrm{LED-}k$ and the BD, respectively.
The orientation of the BD, which is given by $\alpha_{k} = \phi_{k}-\psi_{k}$, represents the angular deviation of the direction where the axis of the acceptance cone of the BD-equipped photodetector is pointing with respect to the upward direction.
Moreover, the radius of the 2D coverage area of $\textrm{LED-}k$~\cite{alexis2024planning} can be written by
$r_{k}=(h_\textrm{LED}-h_{\textrm{BD}})\tan{(\Psi+\alpha_{k})}$, where $h_\textrm{LED}$ and $h_\textrm{BD}$ are the heights of the LED and the BD, respectively.

The BD-equipped photodetector outputs electrical current 
\begin{equation}
\begin{split}
    i_{\textrm{PD}}(t) &= \eta_{\textrm{O-E}} \left( 
    \Sigma_{i=1}^{K} P_{i}^{\textrm{LED}} H_{i}^{\textrm{VLC}} \right) + n(t)\\
    &= I_{\textrm{AC}}(t) + I_{\textrm{DC}}(t) + n(t),
\end{split}
\end{equation}
where $\eta_{\textrm{O-E}}$ is the photodetector responsivity, $I_{\textrm{AC}}(t)$ and $I_{\textrm{DC}}(t)$ are the AC and DC components, and $n(t)$ is additive white Gaussian noise with variance $N_{0}B$, with $N_{0}$ being the noise power spectral density and $B$ the system operational bandwidth. 
Furthermore, the output AC component carrying the VLC message can be expressed by
\mbox{$I_{\textrm{AC}}(t) = \eta_{\textrm{O-E}} \Sigma_{i=1}^{K} \eta_{\textrm{E-O}}  H_{i}^{\textrm{VLC}} m_{i}(t)$}.
For a dedicated VLC link between LED-$k$ and the BD, VLC signals transmitted from other LED APs are treated as co-channel interference. 
Thus, the signal-to-noise-plus-interference ratio (SINR) of the received VLC signal at the input at the BD is given by
\begin{equation}
    \gamma_{k}^{\textrm{SINR}} = \frac{  \eta_{\textrm{O-E}} P_{k}^{\textrm{LED}} H_{k}^{\textrm{VLC}}} {\eta_{\textrm{O-E}} \Sigma_{i=1, i\neq k}^{K} P_{i}^{\textrm{LED}} H_{i}^{\textrm{VLC}}+ N_{0}B}.
\end{equation}
\subsubsection{Energy Harvesting}
To supply the energy that is needed to power up the BD circuitry, e.g., the backscatter modulator, energy harvesting is performed using the DC component output from the BD-equipped photodetector~\cite{letaief2021hybrid}. The amount of energy that can be harvested upon reception of the VLC signal can be written by
\begin{equation}
    E_{\textrm{h}} = \varepsilon I_{\textrm{DC}} V_{\textrm{OC}} = \varepsilon \eta_{\textrm{O-E}} \left(\Sigma_{i=1}^{K} \eta_{\textrm{E-O}}  
    H_{i}^{\textrm{VLC}} I_{\textrm{Bias}} \right) V_{\textrm{OC}},
\end{equation}
where $\varepsilon$ is the fill factor of the photodetector, $V_{\textrm{OC}}$ is the open circuit voltage given by
\begin{equation}
    V_{\textrm{OC}} = V_{\textrm{t}} \ln \left( 1 + \frac{\eta_{\textrm{O-E}} \left( \Sigma_{i=1}^{K} \eta_{\textrm{E-O}} 
    H_{i}^{\textrm{VLC}} I_{\textrm{Bias}} \right)}{I_0} \right),
\end{equation}
where $V_{\textrm{t}}$ is the thermal voltage and $I_{0}$ is the dark saturation current of the BD-equipped photodetector.
\subsection{Backscatter Channel Model (the second hop)}
The BC link involves an RFS, a BD, and UEs equipped with RF receivers. 
The BD modulates the carriers emitted by the RFS using photovoltaic-converted VLC signals, and subsequently, backscatters the modulated signal to the UEs.
The received backscatter signal at the UE can then be expressed as~\cite{griffin2009linkbudget}
\begin{equation}
    y(t) = \sqrt{\xi G_{\textrm{T}} G_{\textrm{R}}} G_{\textrm{BD}} h_{\textrm{f}}(t)  h_{\textrm{b}}(t) c(t) I_{\textrm{AC}}(t) + n(t),
\end{equation}
\noindent where $\xi$ denotes the backscatter efficiency of the BD, and $G_{\textrm{T}}$, $G_{\textrm{R}}$, and $G_{\textrm{BD}}$ are the antenna gains of the RFS, UE, and BD, respectively.
The symbol $h_{\textrm{f}}$ is the forward channel gain (RFS$\rightarrow$BD) and $h_{\textrm{b}}$ is the backscatter channel gain (BD$\rightarrow$UE) between the RFS and BD, and between the BD and UE, respectively.
The ambient RF signal $c(t)$ coming from the RFS has a power of $ P_{\textrm{c}} = \mathbb{E} \{ \left| c(t) \right|^2 \}$. 
Furthermore, $I_{\textrm{AC}}(t)$ represents the AC component of the photovoltaic-converted signal that is used to modulate the RF carriers.
The term $n(t)$ is additive white Gaussian noise with variance $N_{0} B$.
Moreover, $\xi$ is related to the polarization mismatch $\chi_\textrm{f}$ between the RFS and BD, and $\chi_\textrm{b}$ models the mismatch between the BD and UE. 
Finally, the modulation factor of the BD is given by $M$, and $\Theta$ represents an object penalty, and thus the backscatter efficiency can be expressed by 
$\xi = (\chi_\textrm{f} \chi_\textrm{b} M)/ \Theta^{2}$~\cite{griffin2009linkbudget}.

For practical purposes, the radio propagation of the forward and backscatter channels is described using the 3GPP Indoor Hotspot model~\cite{3gpp38901}. 
The model defines path loss for LoS and non-line-of-sight (NLoS) conditions using~(\ref{eq:3GPPInHPathLoss_LoS}) and~(\ref{eq:3GPPInHPathLoss_NLoS}), respectively.
The path losses are expressed as
\begin{equation}
\begin{split}
\mathcal{L}_{\textrm{LoS,dB}} = 32.4 + 17.3\log_{10}(d_{\textrm{3D}})  + 20\log_{10}(f_{\textrm{c}}) + \zeta_\textrm{LoS,dB},
\end{split}
  \label{eq:3GPPInHPathLoss_LoS}
\end{equation}
\begin{equation}
\begin{split}
\mathcal{L}_{\textrm{NLoS,dB}} = \max(\mathcal{L}_{\textrm{LoS,dB}}, 17.3 + 38.3\log_{10}(d_{\textrm{3D}}) \\ + 24.9\log_{10}(f_{\textrm{c}}) + \zeta_\textrm{NLoS,dB}).
\end{split}
  \label{eq:3GPPInHPathLoss_NLoS}
\end{equation}
In~\eqref{eq:3GPPInHPathLoss_LoS} and~\eqref{eq:3GPPInHPathLoss_NLoS}, $d_{\textrm{3D}}$ represents the 3D Euclidean distance, $f_{\textrm{c}}$ is the carrier frequency in gigahertz, and $\zeta_\textrm{LoS,dB}$ is the shadowing factor modeled as a lognormal random variable with $\sigma_{\textrm{LoS}} = 3$ for  \mbox{$1 \leq d_{\textrm{3D}} \leq 150$~m}.
Similarly, $\zeta_\textrm{NLoS,dB}$ is the shadowing factor with $\sigma_{\textrm{NLoS}} = 8.03$ for \mbox{$1 \leq d_{\textrm{3D}} \leq 150$~m}.
The probability of a LoS connection varies with the indoor environment. In an open indoor environment with few obstructions, the LoS probability is given by~\cite{3gpp38901}
\begin{equation} \label{eq:Indoor3GPP_Pr_LoS_Open}
\textrm{Pr}_{\textrm{LoS}}^{\textrm{open}} =
\begin{cases}
1, & d_{\textrm{2D}} \leq 5~\textrm{m}, \\
e^{-\frac{d_{\textrm{2D}} - 5}{70.8}}, & 5 < d_{\textrm{2D}} \leq 49~\textrm{m}, \\
0.54 e^{-\frac{d_{\textrm{2D}} - 49}{211.7}}, & d_{\textrm{2D}} > 49~\textrm{m},
\end{cases}
\end{equation}
where $d_{\textrm{2D}}$ is the 2D Euclidean distance.
In a mixed indoor environment with denser obstacles and more complex multipath effects, the LoS probability is given by~\cite{3gpp38901}
\begin{equation} \label{eq:Indoor3GPP_Pr_LoS_Mix}
\textrm{Pr}_{\textrm{LoS}}^{\textrm{mix}} =
\begin{cases}
1, & d_{\textrm{2D}} \leq 1.2~\textrm{m}, \\
e^{-\frac{d_{\textrm{2D}} - 1.2}{4.7}}, & 1.2 < d_{\textrm{2D}} \leq 6.5~\textrm{m}, \\
0.32 e^{-\frac{d_{\textrm{2D}} - 6.5}{32.6}}, & d_{\textrm{2D}} > 6.5~\textrm{m}.
\end{cases}
\end{equation}
Finally, the expected value of the overall path loss at a given distance, which combines the LoS and NLoS conditions weighted by their respective probabilities, is expressed as
\begin{equation}
\mathcal{L}_{\textrm{dB}} = \textrm{Pr}_{\textrm{LoS}}\cdot \mathcal{L}_{\textrm{LoS,dB}} + (1-\textrm{Pr}_{\textrm{LoS}})\cdot \mathcal{L}_{\textrm{NLoS,dB}}.
  \label{eq:3GPPInHPathLoss_Overall}
\end{equation}
Hence, the signal-to-noise ratio (SNR) of the UE-received backscatter signal can be expressed as
\begin{equation}
\label{eqn:bc_snr_2}
    \gamma_{\textrm{BC}}^{\textrm{SNR}} = \frac{\xi G_{\textrm{T}} G_{\textrm{R}} G_{\textrm{BD}}^2 I^2_{\textrm{AC}} P_{\textrm{c}} \mathcal{L}_{\textrm{f}}^{-1} \mathcal{L}_{\textrm{b}}^{-1}}{N_{0} B},
\end{equation}
where $\mathcal{L}_{\textrm{f}} = |h_{\textrm{f}}|^{-2} $ and $ \mathcal{L}_{\textrm{b}} = |h_{\textrm{b}}|^{-2}$ represent the path loss of the forward channel and backscatter channel, respectively.

\section{Outage Performance Analysis of Joint VLC-BC Relaying System in FBL Regime}
This section analyzes the outage performance of the proposed system in the FBL regime, thereby evaluating the system behavior under short-packet traffic constraints common in IoT applications.
Conventional information-theoretic approaches typically assume infinitely long codewords to approach Shannon capacity. 
However, since most of the IoT communication systems under consideration involve sporadic transmissions of dozens or a few hundred bits, the so-called infinite blocklength approach does not work properly to assess the end-to-end performance. 
As a result, classical asymptotic metrics, such as ergodic or outage capacity, are less meaningful in practice. 
Hence, FBL theory~\cite{verdu2010fbl} provides a more accurate characterization of the trade-off between achievable data rate, latency, and reliability for short-packet traffic, making it well-suited for the proposed hybrid VLC-RF relaying scheme with a low-data-rate requirement. 
In the FBL regime, it is possible to capture the practical effects of blocklength, code rate, and target error probability on achievable data rate and outage of the system.

\subsection{Achievable Rate and Outage Probability of VLC Link}
For the VLC link between the dedicated LED AP and the BD, a lower bound on the channel capacity~\cite{alexis2024planning} is given by
\mbox{$R^{\textrm{VLC}} \;=\; B\,\log_{2}\!\left(1 + \tfrac{e}{2\pi}\,\gamma_{\textrm{VLC}}^{\textrm{SINR}}\right)$}, where $B$ is the system communication bandwidth. 
In the FBL regime, we define the effective SINR as $\gamma' = \tfrac{e}{2\pi}\,\gamma_{\textrm{VLC}}^{\textrm{SINR}}$, leading to a channel dispersion term
\begin{equation}
    V^{\textrm{VLC}}(\gamma') \;=\; \frac{2\,\gamma'}{1+\gamma'}\,(\log_{2}e)^2.
\end{equation}
Hence, the achievable data rate of the VLC link with FBL is approximated by
\begin{align}
    R_{\textrm{FBL}}^{\textrm{VLC}} 
    &\;=\; B\,\log_{2}\left(1+\gamma'\right)
     \;-\; \sqrt{\tfrac{V^{\textrm{VLC}}(\gamma')}{u}}\;\,Q^{-1}(\epsilon),
    \label{eq:vlc_fbl_sectionIII}
\end{align}
where $u$ denotes the blocklength, i.e., the number of channel uses, $\epsilon$ is the target error probability, and $Q^{-1}(\cdot)$ denotes the inverse of the Gaussian $Q$-function. 
The code rate is defined by $R_{\textrm{c}}=\frac{\textrm{packet size}}{u}$ measured in bits per channel use, where a lower $R_{\textrm{c}}$ indicates more redundant codes.
An outage event for the VLC link occurs if $R_{\textrm{FBL}}^{\textrm{VLC}}$ falls below the system desired data rate threshold $R_{\textrm{th}}$, and its probability is expressed by
\begin{equation}
    P_{\textrm{VLC}}^{\mathcal{O}} = P(R_{\textrm{FBL}}^{\textrm{VLC}}  < R_{\textrm{th}}).
\end{equation}

\subsection{Achievable Rate and Outage Probability of BC Link}
For the BC link, Shannon capacity is given by
\mbox{$R^{\textrm{BC}}(t) = B \log_2 \left( 1 + \gamma_{\textrm{BC}}^{\textrm{SNR}} \right)$}.
In the FBL regime with AWGN, the channel dispersion of the backscatter link can be written as
\begin{equation}
    V^{\textrm{BC}}(\gamma_{\textrm{BC}}^{\textrm{SNR}}) = \left(1 - \frac{1}{(1+\gamma_{\textrm{BC}}^{\textrm{SNR}})^2}\right)\,\log_2^2(e).
\end{equation}
Thus, the achievable data rate for the backscatter link is approximated by
\begin{equation}
    R_{\textrm{FBL}}^{\textrm{BC}} 
    \;=\; \log_{2}\left(1 + \gamma_{\textrm{BC}}^{\textrm{SNR}}\right)
    \;-\; \sqrt{\tfrac{V^{\textrm{BC}}(\gamma_{\textrm{BC}}^{\textrm{SNR}})}{u}}\;\,Q^{-1}(\epsilon).
    \label{eq:bc_fbl_sectionIII}
\end{equation}
Similarly, an outage on the BC link occurs with a probability expressed by
\begin{equation}
    P_{\textrm{BC}}^{\mathcal{O}} = P(R_{\textrm{FBL}}^{\textrm{BC}}  < R_{\textrm{th}}).
\end{equation}

\subsection{Overall Outage Probability}
Since the system is declared in outage if either the VLC link or the BC link fails to meet the target rate $R_{\textrm{th}}$, the overall outage probability can be written as
\begin{equation}
    P_{\textrm{VLC-BC}}^{\mathcal{O}} 
    \;=\; 1 \;-\; P\!\left\{ R^{\textrm{VLC}}_{\textrm{FBL}} \!\ge\! R_{\textrm{th}}\right\}\,
          P\!\left\{ R^{\textrm{BC}}_{\textrm{FBL}} \!\ge\! R_{\textrm{th}}\right\},
    \label{eq:pout_sectionIII}
\end{equation}
assuming that the two links are independent~\cite{raouf2024outage},
which is equivalent to
\mbox{$     P_{\textrm{VLC-BC}}^{\mathcal{O}} \;=\; P_{\textrm{VLC}}^{\mathcal{O}} \;+\; P_{\textrm{BC}}^{\mathcal{O}}
     \;-\; P_{\textrm{VLC}}^{\mathcal{O}}\;P_{\textrm{BC}}^{\mathcal{O}}
$}.

\section{Numerical Results}
    \begin{table}[!t]
    \centering
    \caption{Simulation Parameters}
    \resizebox{\columnwidth}{!}{%
    \begin{tabular}{l|l|l}
        \hline
        \textbf{Parameter}          & \textbf{Value}    & \textbf{Description}\\ \hline
        $W \times L$            & $10 \times 10$        & Default size of target area [m$^2$]        \\
        $h_{\text{LED}}, h_{\text{BD}}, h_{\text{RFS}}, h_{\text{UE}}$ & 2.5, 1.5, 3.0, 1.5        & Height of LED APs, BD, RFS, and UE [m]\\
        $(x,y)_{\textrm{LED}}$             & \makecell{(2, 2) (5, 2) (8, 2)\\ (2, 5) (5, 5) (8, 5)\\ (2, 8) (5, 8) (8, 8)}& 2D positions of LED APs [m]\\
        $(x,y)_{\textrm{RFS}}$             & (5, 5)          & 2D position of RFS [m]\\
        \hline
        $A_{\text{PD}}$             & 0.05      & Active area of the BD-equipped photovoltaic cell [m$^2$]\\
        $\Phi_{\text{max}}$ & 60     & LED radiation semi-angle at half power [deg]\\
        $\Psi$              & 60     & FoV semi-angle of the PD’s light acceptance cone [deg]\\
        $\eta_{\textrm{E-O}}$  & 20  & LED power electric-optical conversion factor [W/A]\\
        $\eta_{\textrm{O-E}}$  & 0.5 & Responsivity of the BD-equipped photodetector   [A/W]\\
        $B$                 & 50    & System operational bandwidth [kHz]\\
        $\varepsilon$       & 0.75  & Fill factor\\
        $I_{\text{Bias}}$    & 0.75     & LED driving current [A]\\
        $I_{0}$             & $10^{-9}$   & Dark saturation current [A]\\
        $V_{\text{t}}$     & $25\times 10^{-3}$   & Thermal voltage [V]\\
        \hline
        $P_{\text{c}}$          & 23                     & RFS carrier power [dBm]\\
        $f_{\text{c}}$          & 2.45                     & RFS carrier frequency [GHz]\\
        $G_{\text{T}}$, $G_{\text{R}}$,  $G_{\text{BD}}$     & 8, 3, 1.5  & Antenna gains of RFS, UE, and BD [dBi]\\
        $\chi_{\textrm{f}}, \chi_{\textrm{b}}$           & 0.5, 0.5  & Polarization mismatch\\
        $M$              & 0.5               & Modulation factor\\
        $\Theta$         & 0                  & On-object penalty [dB]\\
        $u$              & 64                 & Channel uses, i.e., blocklength\\
        $R_{\textrm{c}}$ & 3/4                & Code rate\\
        $\epsilon$ & $10^{-3}$                & Target error probability\\
        $R_{\textrm{th}}$ & 10         & System desired data rate threshold [kbit/s]\\
        $N_0$            & -174               & Noise power spectral density [dBm/Hz]\\
        $(\sigma_{\text{LoS}},\sigma_{\text{NLoS}})$ & (3, 8.03) & Shadowing factors in the 3GPP model [dB]\\
        \hline
    \end{tabular}}
    \label{tab:parameters}
    \vspace{-10pt}
    \end{table}
This section presents numerical results that illustrate the outage performance and achievable data rate of the joint VLC-BC relaying scheme.
Simulation parameters are as follows: 9 LED APs are deployed in an indoor space measuring $10\times10~\textrm{m}^2$, with positions detailed in Table~\ref{tab:parameters}. LED APs transmit VLC signals via TDM.
An RFS is deployed at the center of the 2D area on the ceiling, emitting a 2.45~GHz carrier wave. 
The position of the BD is randomly changed within the VLC coverage area of a dedicated LED AP, following a \emph{uniform} distribution, while signals from other LED APs are treated as light interference.
Similarly, the position of the UE is randomly changed within the indoor space according to a \emph{uniform} distribution.
A summary of parameters that were used to obtain the simulation results is listed in Table~\ref{tab:parameters}.
A total of $10^5$ Monte Carlo simulations have been conducted to assess the overall link outage performance and achievable data rate in the FBL regime, considering significant variables such as the height and orientation of the BD, code rate of the end-to-end link, and indoor radio propagation conditions.
It is noteworthy that the system operates with a limited bandwidth and low-data-rate communications within the scope of Ambient IoT~\cite{3gpp38848}.

\begin{figure}
\centering
\includegraphics[width=0.98\columnwidth]{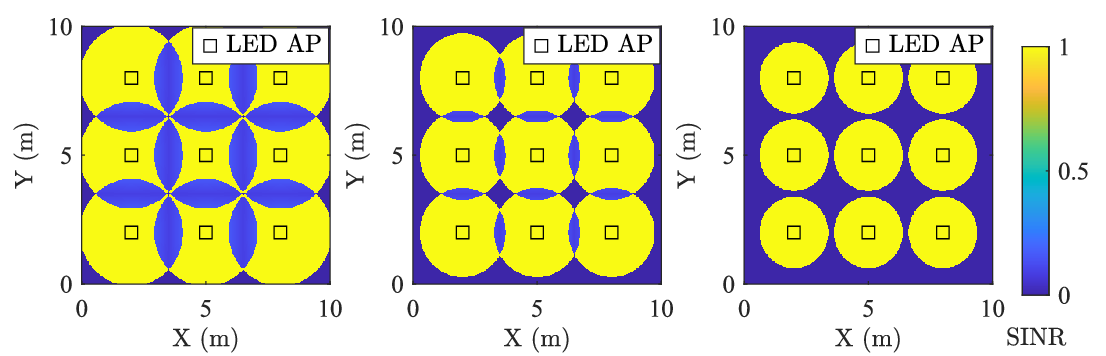}
\caption{Normalized SINR of the BD-received VLC signal with different BD heights: 1.3, 1.5, and 1.7~m (from left to right).}
\label{fig:sinr}
\vspace{-10pt}
\end{figure}
\begin{figure}
\centering
\includegraphics[width=0.98\columnwidth]{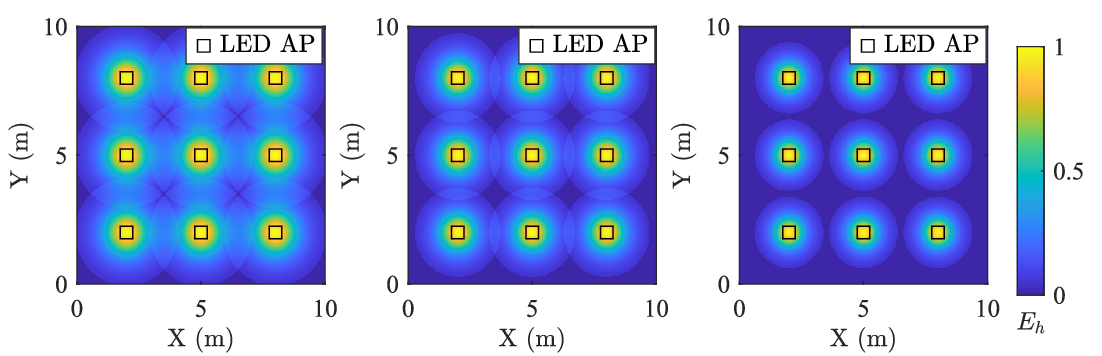}
\caption{Normalized BD-harvested energy from LED APs with different BD heights: 1.3, 1.5, and 1.7~m (from left to right).}
\label{fig:eh}
\vspace{-10pt}
\end{figure}
Fig.~\ref{fig:sinr} shows the SINR heatmap for the VLC signals received by the BD across the space for various BD heights.
The color scheme, ranging from dark to bright, represents the increase in SINR levels.
The circles indicate the VLC coverage areas, with each LED AP having its own dedicated coverage region.
The overlapping shaded areas between the circles indicate regions suffering from strong co-channel interference emitted by adjacent VLC coverage.
As the height of the BD increases, the area affected by light interference decreases. However, the overall VLC coverage area also decreases, limiting the available reception area.
Furthermore, Fig.~\ref{fig:eh} shows the heatmap of energy harvested by the BD from received VLC signals across the space for various BD heights.
In this scenario, while operating within its dedicated VLC coverage, the BD also harvests energy from interfering light emitted by adjacent LED APs.
As the vertical position of the BD increases, the harvestable energy from the dedicated LED AP increases due to the reduced distance between them; however, the harvestable energy from adjacent LED APs decreases as a result of diminished interfering signal power. 
%
\begin{figure}
\centering
\includegraphics[width=0.6\columnwidth]{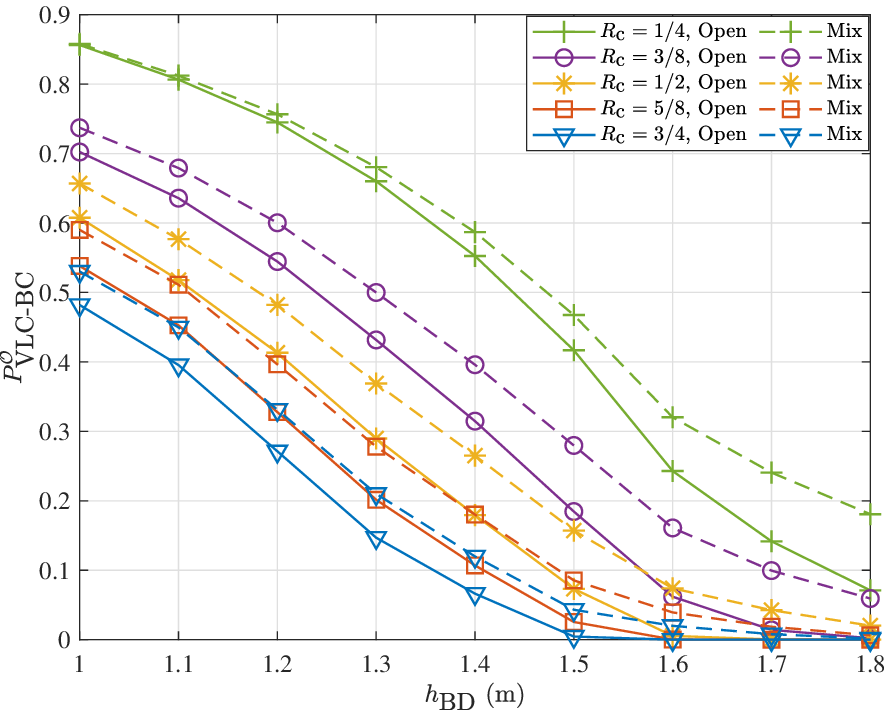}
\caption{Overall outage probability versus BD heights with varying code rates.}
\label{fig:out_hbd_codeRate}
\end{figure}
\begin{figure}
\centering
\includegraphics[width=0.6\columnwidth]{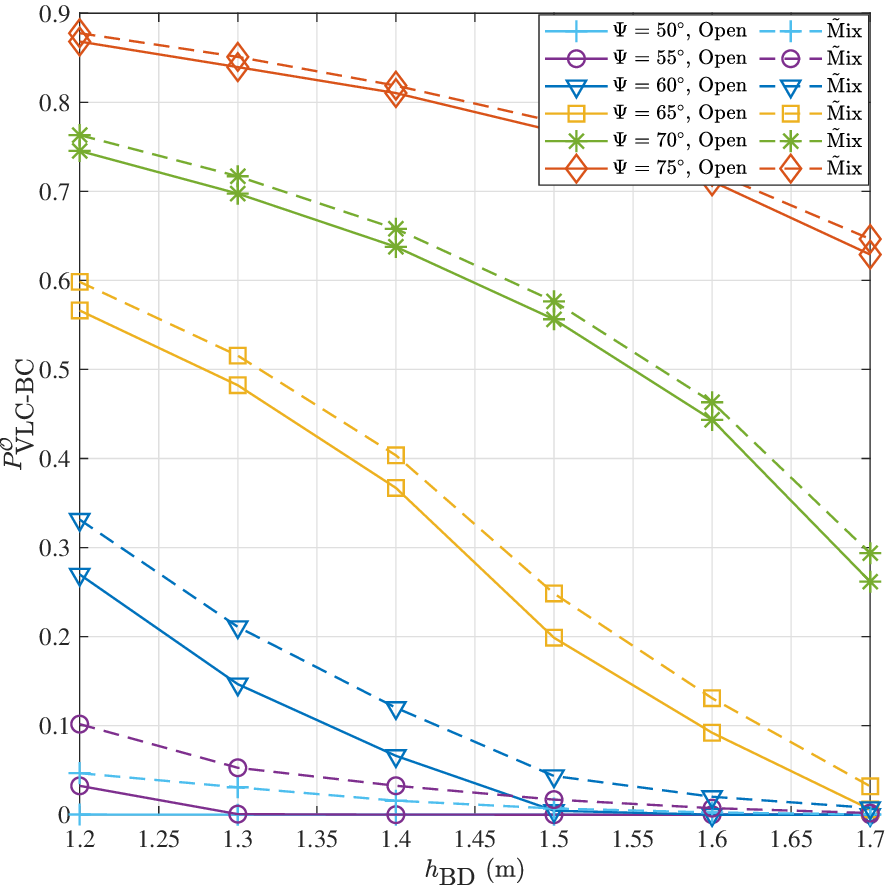}
\vspace{-10pt}
\caption{Overall outage probability versus different BD heights with varying FoV semi-angle of BD-equipped photodetector.}
\label{fig:out_hbd_fov}
\end{figure}

Fig.~\ref{fig:out_hbd_codeRate} depicts the overall outage probability $P_{\textrm{VLC-BC}}^{\mathcal{O}}$ for various heights of the BD considering different code rates in the FBL regime.
The outage probability decreases as the vertical position of the BD increases, due to reduced reception of interfering light and consequently higher SINR in the VLC link.
With a fixed data rate threshold $R_{\textrm{th}}=10$~kbps for the end-to-end link, the use of a higher code rate $R_{\textrm{c}}$ with less redundancy results in a lower outage probability by improving the effective data rate.
Furthermore, the presented result distinguishes between open-indoor and mixed-indoor radio propagation conditions for the BC link, represented by solid and dashed lines, respectively.
Outage performance in an open-indoor environment consistently outperforms that in a mixed-indoor environment, as the former offers a higher LoS probability for the BC link.
Furthermore, Fig.~\ref{fig:out_hbd_fov} presents the overall outage probability for various BD heights under different FoV semi-angles of the BD-equipped photodetector.
Similarly, the outage decreases as the vertical position of the BD increases, while a smaller FoV leads to a lower outage probability by capturing less interfering light.

\begin{figure}
\centering
\includegraphics[width=0.6\columnwidth]{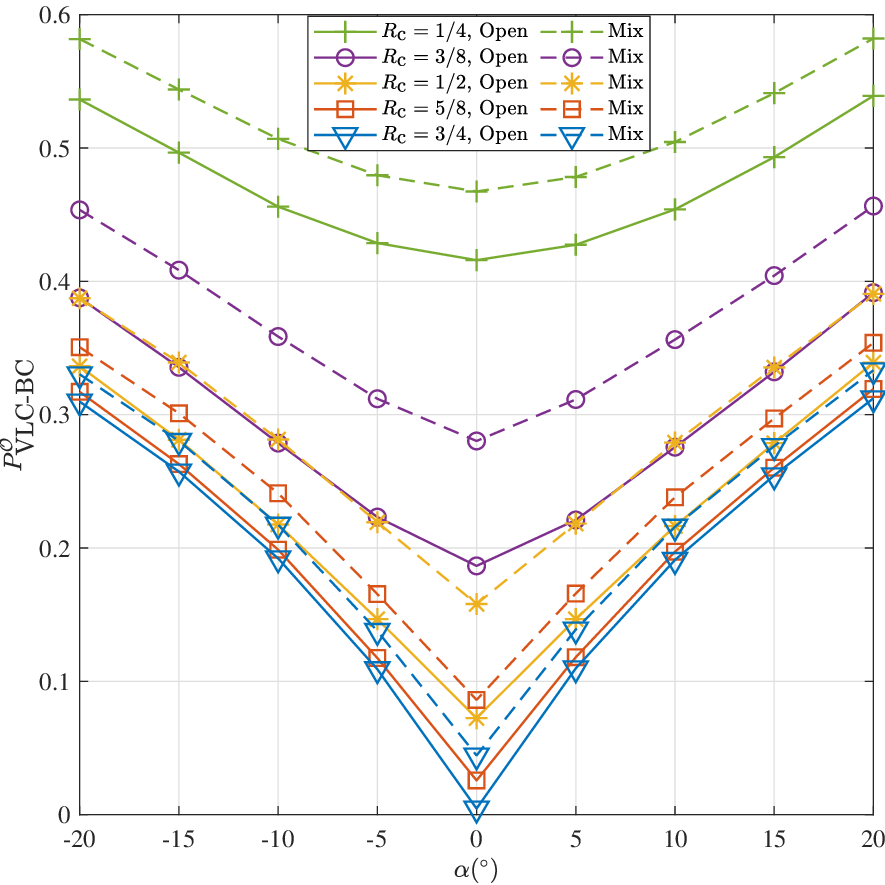}
\caption{Overall outage probability versus BD orientations with varying code rates.}
\label{fig:out_ori_codeRate}
\vspace{-15pt}
\end{figure}
Fig.~\ref{fig:out_ori_codeRate} shows the overall outage probability for various BD orientations at different code rates in the FBL regime.
The lowest outage occurs when the BD is oriented vertically upward toward the dedicated LED AP ($\alpha=0^{\circ}$). 
The same effect of the code rate can also be observed where a higher $R_{\textrm{c}}$ contributes to a lower outage.
Moreover, better outage performance is observed with open-indoor backscatter propagation.

\begin{figure}
\centering
\includegraphics[width=0.6\columnwidth]{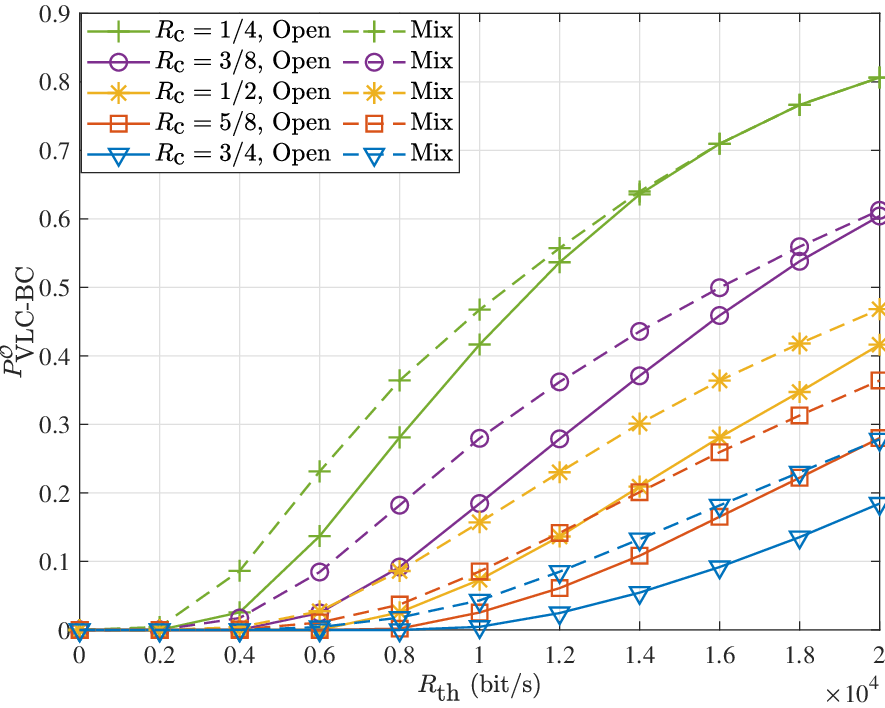}
\vspace{-10pt}
\caption{Overall outage probability versus data rate threshold requirements, with varying code rates.}
\label{fig:out_rth_codeRate}
\end{figure}
\begin{figure}
\centering
\includegraphics[width=0.6\columnwidth]{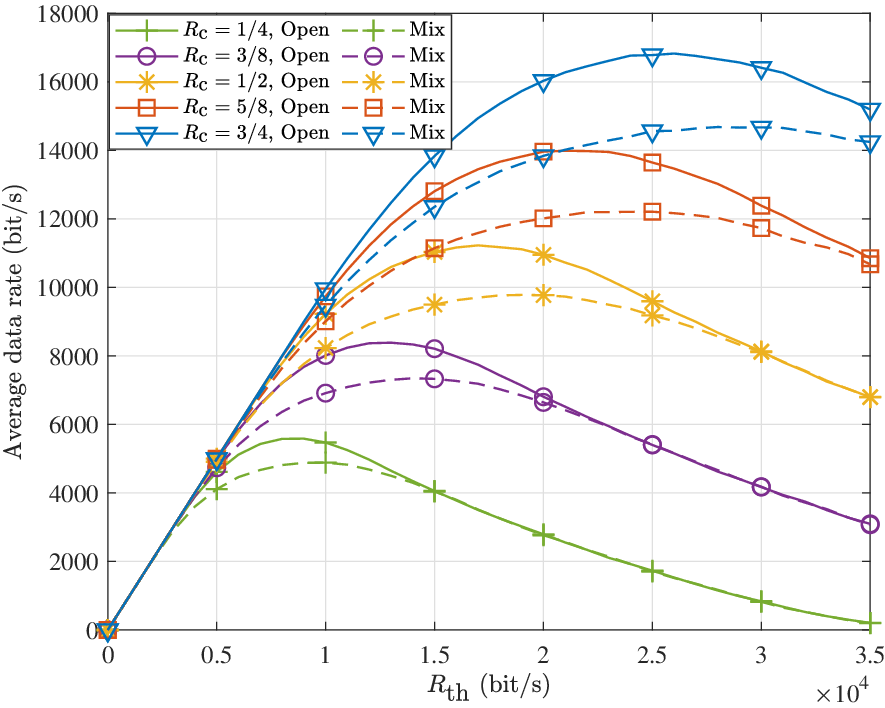}
\vspace{-10pt}
\caption{Average data rate of the system versus data rate threshold requirements, with varying code rates.}
\label{fig:tp_rth_codeRate}
\end{figure}
\begin{figure}[!ht]
\centering
\includegraphics[width=0.6\columnwidth]{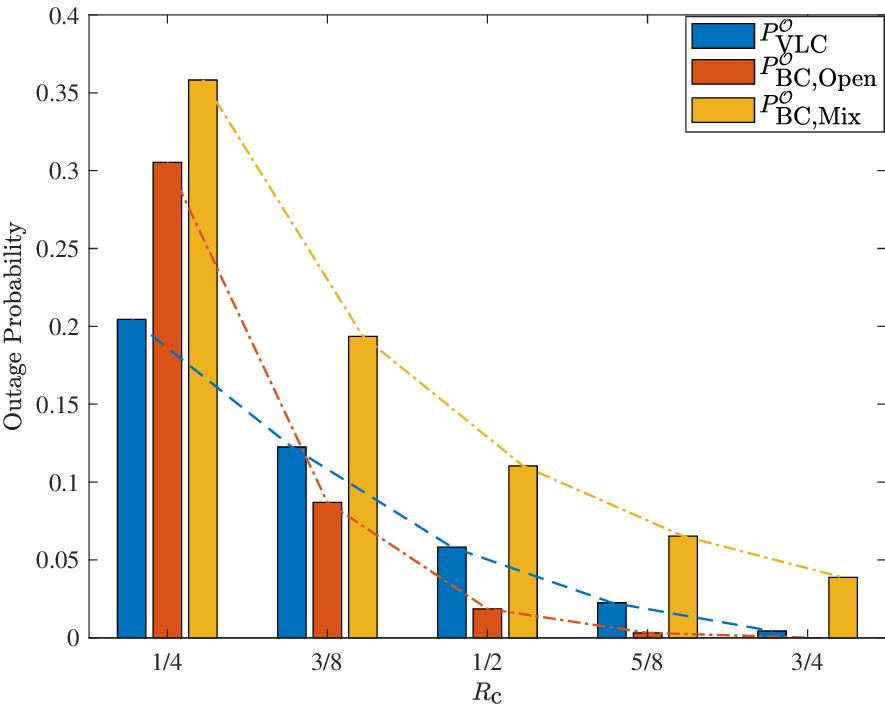}
\vspace{-10pt}
\caption{Outage probability of VLC and BC links versus code rates.}
\label{fig:out_codeRate_hist}
\vspace{-15pt}
\end{figure}
Fig.~\ref{fig:out_rth_codeRate} presents the overall outage probability for various data rate requirements $R_{\textrm{th}}$ of the system, under different code rates in the FBL regime.
The outage probability increases as the data rate requirement increases. Furthermore, a lower code rate, which introduces more redundancy and reduces effective data rate, leads to an increased outage probability.
Fig.~\ref{fig:tp_rth_codeRate} presents the average end-to-end data rate, calculated as 
\mbox{$(1-P_{\textrm{VLC-BC}}^{\mathcal{O}}) R_{\textrm{th}}$}, for various data rate requirements of the system at different code rates.
The average data rate initially increases with $R_{\textrm{th}}$ and subsequently decreases due to system limitations.
Similarly, a higher code rate yields a higher average data rate compared to a lower code rate that introduces more redundancy.
Moreover, the outage performance is better when the BC link is under the open-indoor propagation condition.
Fig.~\ref{fig:out_codeRate_hist} illustrates the separate outage probabilities for the VLC link $P_{\textrm{VLC}}^{\mathcal{O}}$, and BC link $P_{\textrm{BC}}^{\mathcal{O}}$ across different code rates. Both open- and mixed-indoor radio propagation models are implemented for the BC link.
The chart indicates that a higher code rate leads to a lower outage probability for both the VLC and BC links.

\section{Conclusion}
This paper has presented a joint VLC-BC relaying framework under the FBL constraint that addresses the challenge of energy-efficient and low-complexity IoT communications. 
The proposed system leverages LED APs to deliver VLC signals, which are received by a BD that harvests optical energy and relays information to UEs via ambient RF carriers.
Simulation results have revealed that careful configuration of the placement and orientation of the BD, as well as the selection of code rate of the system, can enhance both reliability and achievable data rate. Integrating this approach into hybrid VLC/RF networks promises to support energy-efficient IoT with broad compatibility with conventional RF receivers, offering a pathway for energy-neutral A-IoT solutions.
\vspace{-0pt}
\bibliographystyle{IEEEtran}
\bibliography{reference}

\end{document}